\documentclass[10pt]{iopart}
\usepackage{graphicx}
\usepackage{xcolor}  
\usepackage{hyperref}
\usepackage[utf8]{inputenc} 
\begin{document}

\title{Voltage- and temperature-dependent rare-earth dopant contribution to the interfacial magnetic anisotropy}
\author{Alejandro O. Leon$^{1}$ and Gerrit E. W. Bauer$^{2,3,4}$}

\begin{abstract}
The control of magnetic materials and devices by voltages without electric currents holds the promise of power-saving nano-scale devices. Here we study the temperature-dependent voltage control of the magnetic anisotropy caused by rare-earth (RE) local moments at an interface between a magnetic metal and a non-magnetic insulator, such as Co$|$(RE)$|$MgO. Based on a Stevens operator representation of crystal and applied field effects, we find large dominantly quadrupolar intrinsic and field-induced interface anisotropies at room temperature. We suggest improved functionalities of transition metal tunnel junctions by dusting their interfaces with rare earths.
\end{abstract}

\address{$^1$Departamento de F\'isica, Facultad de Ciencias Naturales, 
Matem\'atica y del Medio Ambiente,
Universidad Tecnol\'ogica Metropolitana, Las Palmeras
3360, Ñuñoa 780-0003, Santiago, Chile}
\address{$^2$ WPI-AIMR, Tohoku University, Sendai 980-8577, Japan}
\address{$^3$ Institute for Materials
Research $\&$ CSRN, Tohoku University, Sendai 980-8577, Japan}
\address{$^4$ Zernike
Institute for Advanced Materials, Groningen University, The Netherlands}
\ead{aleonv@utem.cl}

\vspace{10pt} \begin{indented}
\item[]March 2020
\end{indented}

\submitto{\JPCM} \ioptwocol

\section{Introduction}

The magnetic order can be excited by magnetic fields, spin~\cite{STT,TmIG0}
and heat~\cite{SpinCaloritronics,OtaniSpinConv} currents, mechanical rotations
and sound waves~\cite{MechanicalWaves1,MechanicalWaves2}, optical fields in
cavities~\cite{Cavitronics1,Cavitronics2}, and electric
fields~\cite{Suzuki2011,Nozaki2012,ParametricVCMA,GeneralReferenceEField,GeneralReferenceEField2,VCMARecentReview}. A mechanism of the latter is
\textit{voltage-control of magnetic anisotropy} (VCMA), which avoids electric
currents and thereby Joule heating. A time-dependent applied electric field
can assist or fully actuate magnetization switching~\cite{GeneralReferenceEField,GeneralReferenceEField2,VCMARecentReview,Shiota2009, Assisted2,Assisted,Kanai2012,SwitchingNp1,SwitchingNp2}, and excite the
ferromagnetic resonance~\cite{Nozaki2012,GeneralReferenceEField,GeneralReferenceEField2,VCMARecentReview,Zhu2012}. However, in order to become useful, the VCMA should be enhanced. This can be realized by, for example, improving
interface properties~\cite{Interfaces2,VCMARecentReview}, thermal
stability~\cite{ThermalStability}, employing higher-order magnetic
anisotropies~\cite{VCMAHigher1}, and reducing temperature
dependences~\cite{VCMAHigher2}. The control of magnetic properties by electric
fields has also been demonstrated or proposed in magnetoelectric
materials~\cite{Gerhard2010,Yamada2011,Sekine2016}, by proximity
effects~\cite{Platinum,PtProximityMag,ChibaTanaka}, by nuclear spin resonance in
single-molecule magnets~\cite{SMMWernsdorfer}, and by the tuning of exchange
interactions~\cite{You2005,Haney2009a,Tang2009,New2017,RKKYMultilayers1,RKKYImpurities}.

The electrostatic environment of a local moment affects its magnetic energy
via the spin-orbit interaction (SOI)~\cite{BookSkomski1,BookSkomski2}. In
transition-metal atoms such as Fe, Co, and Ni with partially filled 3d
subshells, the electrostatic interaction with neighboring atoms, $E_{CF}\sim
1$~eV is much larger than the SOI $E_{SOI}\sim0.05$~eV~\cite{BookSkomski1},
which implies that the orbital momentum of transition-metal ions is easily
quenched, while the relatively large 3d orbital radius favors band formation and
itinerant magnetism. The opposite occurs for the lanthanide series, i.e.,
atoms from lanthanum (La with atomic number 57) to lutetium (Lu with atomic
number 71). The \textit{rare earths} (RE) also include non-magnetic
scandium (Sc) and yttrium (Y). The ground states of the lanthanide La$^{+3}$, Eu$^{+3}$, and Lu$^{+3}$ ions are also not magnetic. The half-filled subshell of the magnetic ion Gd$^{+3}$
lacks orbital angular momentum and, therefore, SOI. Except for La$^{+3}$, Eu$^{+3}$, Gd$^{+3}$, and Lu$^{+3}$,
the 4f SOI energy of the lanthanide series $E_{SOI}\sim0.2$ eV is much stronger than crystal-field energies
$E_{CF}\sim0.01$ eV~\cite{BookSkomski1}, so their orbital momenta are
atomic-like and not quenched. The magnetism of lanthanide-containing compounds
can be understood by models that proceed from an atomic picture. Nevertheless,
since the crystal fields lock to their spin-orbit induced anisotropic charge
distributions, large magnetocrystalline anisotropies can be achieved.

The mechanism for the VCMA of RE moments is the electric
field-induced torque on an anisotropic 4f charge distribution with rigidly
coupled magnetic moment by the electric quadrupolar coupling~\cite{OurVoltage}.
This torque is communicated to the magnetic order via the exchange interaction.

Here, we predict that an interfacial RE dusting of transition-metal magnetic tunnel junctions can enhance its VCMA efficiency. We study the temperature dependence of the VCMA of RE moments, as well as the role of higher-order anisotropy constants. The latter issue has been addressed in transition-metal systems~\cite{VCMAHigher1,VCMAHigher2}, where the first- and second-order contributions partially cancel in the total VCMA. We calculate the magnetic anisotropy constants (MACs) of a rare-earth ion in the presence of an electric field, assuming a strong exchange coupling with the system magnetization. The effect is strongest for a RE at an interface between a magnetic metal and a non-magnetic insulator, such as Co$|$MgO. The Hamiltonian of the local moment in an angular momentum basis leads to so-called Stevens operators that can be easily diagonalized. We extract the intrinsic and field-induced MACs from the corresponding temperature-dependent free energy as a function of temperature.

\section{Single-ion magnetic anisotropy}

The 4f atomic radius is small compared to that of other filled atomic shells,
which isolates the 4f electrons from other atoms in
compounds~\cite{BookJensen}. Consequently, the crystal fields that would
quench the orbital momentum of 3d transition metals only slightly affect 4f
electron ground-state configurations. The 4f subshell is characterized by a
spin ($\mathbf{S}$), an orbital momentum ($\mathbf{L}$), and a total angular
momentum ($\mathbf{J}=\mathbf{L}+\mathbf{S}$). In the basis $|S,L,J,J_{z}%
\rangle$,
\begin{eqnarray*}
\mathbf{S}^{2}|S,L,J,J_{z}\rangle & =\hbar^{2}S\left(  S+1\right)
|S,L,J,J_{z}\rangle,\\
\mathbf{L}^{2}|S,L,J,J_{z}\rangle & =\hbar^{2}L\left(  L+1\right)
|S,L,J,J_{z}\rangle,\\
\mathbf{J}^{2}|S,L,J,J_{z}\rangle & =\hbar^{2}J\left(  J+1\right)
|S,L,J,J_{z}\rangle,\\
\hat{J}_{z}|S,L,J,J_{z}\rangle & =\hbar J_{z}|S,L,J,J_{z}\rangle,
\end{eqnarray*}
where $S$ and $L$ are governed by Hund's first and second rules, respectively.
The third rule determines the multiplet $J=L\pm S$, where the $-$ and $+$ is
for the light (i.e., less than half-filled 4f shell with an atomic number less
than 64) and heavy REs, respectively. We list the $S$, $L$, and $J$ for the
whole 4f series in table~\ref{TableHundRules}. In the following, we focus on
the ground-state manifold with constant $S$, $L$, and $J$ numbers. This multiplet of
$J=L\pm S$ has $2J+1$ states that are degenerate in the absence of
electromagnetic fields. Also,
\begin{eqnarray*}
\mathbf{S}  & =(g_{J}-1)\mathbf{J},\\
\mathbf{L}  & =(2-g_{J})\mathbf{J},\\
\mathbf{L}+2\mathbf{S}  & =g_{J}\mathbf{J},
\end{eqnarray*}
where $g_{J}=3/2+[S(S+1)-L(L+1)]/[2J(J+1)]$ is the Land\'{e} $g-$factor. The
projections of $\mathbf{S}$, $\mathbf{L}$, and $\mathbf{L}+2\mathbf{S}$ on
$\mathbf{J}$ for lanthanide atoms manifests itself also in the crystal-field
Hamiltonian, as shown in the next subsection.

\subsection{Stevens operators}

Let us consider a crystal site with a potential that is invariant to rotations around the $z$-axis, which can be expanded as
\begin{eqnarray*}
-eV\left(  \mathbf{r}\right)   & =A_{2}^{(0)}\left(  3z^{2}-r^{2}\right)
+A_{4}^{(0)}\left(  35z^{4}-30r^{2}z^{2}+3r^{4}\right) \nonumber\\
& +A_{6}^{(0)}\left(  231z^{6}-315z^{4}r^{2}+105z^{2}r^{4}-5r^{6}\right)  ,
\end{eqnarray*}
where $A_{l}^{(0)}$ is a \textit{uniaxial crystal-field parameter} associated
to the $Y_{l}^{0}$ spherical harmonic function (see \ref{SecAppendixA}),
usually expressed in units of temperature divided by $a_{0}^{l}$, where
$a_{0}\approx 0.53$ \AA \ is the Bohr radius. For example, for the 4f states of
Nd$_{2}$Fe$_{14}$B~\cite{CoeysBook} $A_{2}^{(0)}=304$\thinspace\textrm{K}%
$/a_{0}^{2}$, $A_{4}^{(0)}=-15\,\mathrm{K}/a_{0}^{4}$, and $A_{6}%
^{(0)}=-2\,\mathrm{K}/a_{0}^{6}$. The crystal-field parameters of the 4f and
4g states of other members of the (RE)$_{2}$Fe$_{14}$B family can be found
in Ref.~\cite{CoeysBook}. \begin{table}[t]
\begin{center}%
\begin{tabular}
[c]{|c|c|c|c|c|c|}\hline
Ion & $4f^{n}$ & S & L & J & g$_{J}$\\\hline
Ce$^{3+}$ & $4f^{1}$ & 1/2 & 3 & 5/2 & 6/7\\\hline
Pr$^{3+}$ & $4f^{2}$ & 1 & 5 & 4 & 4/5\\\hline
Nd$^{3+}$ & $4f^{3}$ & 3/2 & 6 & 9/2 & 8/11\\\hline
Pm$^{3+}$ & $4f^{4}$ & 2 & 6 & 4 & 3/5\\\hline
Sm$^{3+}$ & $4f^{5}$ & 5/2 & 5 & 5/2 & 2/7\\\hline
Eu$^{3+}$ & $4f^{6}$ & 3 & 3 & 0 & -\\\hline
Gd$^{3+}$ & $4f^{7}$ & 7/2 & 0 & 7/2 & 2\\\hline
Tb$^{3+}$ & $4f^{8}$ & 3 & 3 & 6 & 3/2\\\hline
Dy$^{3+}$ & $4f^{9}$ & 5/2 & 5 & 15/2 & 4/3\\\hline
Ho$^{3+}$ & $4f^{10}$ & 2 & 6 & 8 & 5/4\\\hline
Er$^{3+}$ & $4f^{11}$ & 3/2 & 6 & 15/2 & 6/5\\\hline
Tm$^{3+}$ & $4f^{12}$ & 1 & 5 & 6 & 7/6\\\hline
Yb$^{3+}$ & $4f^{13}$ & 1/2 & 3 & 7/2 & 8/7\\\hline
\end{tabular}
\end{center}
\caption{Ground-state manifold of the tri-positive 4f ions. $S$, $L$, and $J$
are the quantum numbers associated with $\mathbf{S}^{2}$, $\mathbf{L}^{2}$,
and $\mathbf{J}^{2}$, respectively. $g_{J}$ is the Land\'{e} g-factor.}%
\label{TableHundRules}%
\end{table}

The electrostatic Hamiltonian of $N_{4f}$ electrons in the subshell Hilbert
space can be expanded into
\begin{eqnarray}
\sum_{j=1}^{N_{4f}}\left(  3\hat{z}_{j}^{2}-\hat{r}_{j}^{2}\right)
=\vartheta_{2}\left\langle r^{2}\right\rangle \hat{O}_{2}^{(0)},\nonumber\\
\sum_{j=1}^{N_{4f}}\left(  35\hat{z}_{j}^{4}-30\hat{r}_{j}^{2}\hat{z}_{j}%
^{2}+3\hat{r}_{j}^{4}\right)  =\vartheta_{4}\left\langle r^{4}\right\rangle
\hat{O}_{4}^{(0)},\nonumber\\
\sum_{j=1}^{N_{4f}}h\left(\hat{r}_{j},\hat{z}_{j} \right)
=\vartheta_{6}\left\langle r^{6}\right\rangle \hat{O}_{6}^{(0)},
\label{Stevens}%
\end{eqnarray}
where $h\left(\hat{r}_{j},\hat{z}_{j} \right)=231\hat{z}_{j}^{6}-315\hat{z}_{j}^{4}\hat{r}%
_{j}^{2}+105\hat{z}_{j}^{2}\hat{r}_{j}^{4}-5\hat{r}_{j}^{6}$ and $\hat{z}_{j}$ and $\hat{r}_{j}$ are the operators of the $z$ and the
radial coordinates of the $j$-th electron, respectively. $\langle r^{l}%
\rangle$ is the mean value of $r^{l}$ calculated for a 4f (atomic) radial wave
function. The projection constants $\vartheta_{l}$ are listed in
Table~\ref{TableThetas}, while \textit{Stevens equivalent operators} are
\begin{eqnarray}
\hbar^{2}\hat{O}_{2}^{(0)}  & =3\hat{J}_{z}^{2}-\mathbf{J}^{2},\\
\hbar^{4}\hat{O}_{4}^{(0)}  & =35\hat{J}_{z}^{4}-30\mathbf{J}^{2}\hat{J}%
_{z}^{2}+25\hbar^{2}\hat{J}_{z}^{2}-6\hbar^{2}\mathbf{J}^{2}+3\mathbf{J}%
^{4},\\
\hbar^{6}\hat{O}_{6}^{(0)}  & =231\hat{J}_{z}^{6}-315\mathbf{J}^{2}\hat{J}%
_{z}^{4}+735\hbar^{2}\hat{J}_{z}^{4}+105\mathbf{J}^{4}\hat{J}_{z}^{2}\\
& -525\hbar^{2}\mathbf{J}^{2}\hat{J}_{z}^{2}+294\hbar^{4}\hat{J}_{z}%
^{2}-5\mathbf{J}^{6}+40\hbar^{2}\mathbf{J}^{4}-60\hbar^{4}\mathbf{J}%
^{2}.\nonumber
\end{eqnarray}
Stevens operators for other symmetries are listed
in~\cite{StevensOp1,StevensOp2,StevensOp3}. The total crystal-field
Hamiltonian reads
\begin{equation}
H_{CF}=-e\sum_{j=1}^{N_{4f}}V\left(  \mathbf{r}_{j}\right)  =\sum
_{l=2,4,6}\vartheta_{l}\left\langle r^{l}\right\rangle A_{l}^{(0)}\hat{O}%
_{l}^{(0)}.
\end{equation}
\begin{table}[t]
\begin{center}%
\begin{tabular}
[c]{|c|c|c|c|c|}\hline
Ion & $4f^{n}$  & $10^2\vartheta_2$ & $10^3\vartheta_4$& $10^4\vartheta_6$\\\hline
Ce$^{3+}$ & $4f^{1}$ & -5.71 & 6.35&0\\\hline
Pr$^{3+}$ & $4f^{2}$  & -2.10 & -0.73&0.61\\\hline
Nd$^{3+}$ & $4f^{3}$ & -0.64 &-0.29 &-0.38\\\hline
Pm$^{3+}$ & $4f^{4}$ &  0.77&0.41 &6.69\\\hline
Sm$^{3+}$ & $4f^{5}$ & 4.13 &2.50 &0\\\hline
Eu$^{3+}$ & $4f^{6}$ & - & -&-\\\hline
Gd$^{3+}$ & $4f^{7}$ & - & -&-\\\hline
Tb$^{3+}$ & $4f^{8}$ & -1.01 &0.12&-0.01\\\hline
Dy$^{3+}$ & $4f^{9}$ & -0.63 & -0.06 &0.01\\\hline
Ho$^{3+}$ & $4f^{10}$ & -0.22& -0.03&-0.01\\\hline
Er$^{3+}$ & $4f^{11}$ &0.25  &0.04&0.02\\\hline
Tm$^{3+}$ & $4f^{12}$ & 1.01 & 0.16&-0.06\\\hline
Yb$^{3+}$ & $4f^{13}$ & 3.17 &-1.73& 1.48\\\hline
\end{tabular}
\end{center}
\caption{Projection constants in the Stevens' operators Eqs. (\ref{Stevens})
\cite{StevensOp2}. The nearly ellipsoidal 4f electron density causes a
hierarchy of projection constants, i.e., most ions obey the scaling
$|\vartheta_{2}|\sim10^{-3}-10^{-2}$, $|\vartheta_{4}|\sim10^{-5}-10^{-3}$, and $|\vartheta
_{6}|\sim10^{-6}-10^{-4}$, and then the quadrupole dominates. Some references use the
notation $\alpha_{J}=\vartheta_{2}$, $\beta_{J}=\vartheta_{4}$, and
$\gamma_{J}=\vartheta_{6}$.}%
\label{TableThetas}%
\end{table}

\subsection{Magnetic anisotropy constants}

In several magnets, the exchange interaction strongly couples the 4f local
moments to the magnetization $\mathbf{m}=\sin\theta\left(  \cos\phi
\mathbf{e_{x}}+\sin\phi\mathbf{e_{y}}\right)  +\cos\theta\mathbf{e_{z}}$,
where $\mathbf{e_{j}}$ is the unit vector along the Cartesian axis $j$. Then,
the Hamiltonian $H$ of a single RE atom reads
\begin{equation}
H=H_{CF}+\frac{J_{ex}\left(  g_{J}-1\right)  f(T)}{\hbar}\mathbf{J}%
\cdot\mathbf{m},\label{EqHamiltonianGeneral}%
\end{equation}
where $J_{ex}>0$ is the exchange constant with units of energy. The exchange
coupling favors the parallel alignment between the magnetization $\mathbf{m}$
and the spin contribution to the 4f moment $-\gamma_{e}\left(  g_{J}-1\right)
\mathbf{J}$, with $-\gamma_{e}$ being the electron gyromagnetic ratio. Note
that the 4f spin $\mathbf{S}$ is antiparallel (parallel) to $\mathbf{J}$ for
the light (heavy) lanthanides because of $g_{J}<1 $ ($g_{J}>1$). $f(T)$
parameterizes the temperature dependence of the system
magnetization~\cite{KuzMin1,KuzMin2}
\begin{equation}
f(T)=\left[  1-s\left(  \frac{T}{T_{C}}\right)  ^{3/2}-\left(  1-s\right)
\left(  \frac{T}{T_{C}}\right)  ^{p}\right]  ^{1/3},\label{EqKuzMin}%
\end{equation}
where $T_{C}$ is the Curie temperature, and $s$ and $p$ (with $p>s$) are
material-dependent parameters. For example, for Co~\cite{KuzMin1},
$T_{C}=1385$ K, $s=0.11$, and $p=5/2$; for Fe, $T_{C}=1044$ K, $s=0.35$, and
$p=4$. Empirical expression~(\ref{EqKuzMin}) describes the temperature
dependence between Bloch's law $1-(s/3)\left(  T/T_{C}\right)  ^{3/2}$ for
$T\rightarrow0$ and the critical scaling $\left(  1-T/T_{C}\right)  ^{1/3}$
for $T\rightarrow T_{C}$. Equation~(\ref{EqKuzMin}) is all we need to know
about the magnetic host.

The Helmholtz free energy~\cite{DefK}
\begin{equation}
F=-\frac{1}{\beta}\ln\left[  \sum_{n=1}^{2J+1}e^{-\beta E_{n}}\right]
,\label{EqFHelmholtz}%
\end{equation}
where $\beta=1/(k_{B}T)$, $k_{B}=8.617\times10^{-5}$ eV/K is Boltzmann's
constant, $T$ is temperature, and $E_{n}$ is the $n$-th eigenvalue of
Eq.~(\ref{EqHamiltonianGeneral}). The uniaxial anisotropy energy density can be expanded in the magnetization direction \(\theta\)  as~\cite{BookSkomski1,BookSkomski2}
\begin{equation}
n_{RE}F=K_{1}\sin^{2}\theta+K_{2}\sin^{4}\theta+K_{3}\sin^{6}\theta,
\end{equation}
where $n_{RE}$ is the (surface or volume) density of RE moments and
\begin{eqnarray}
K_{1}  & =\frac{n_{RE}}{2}\lim_{\theta\rightarrow0}\left[  \left(
\partial_{\theta}\right)  ^{2}F\right]  ,\label{EqDefK1}\\
K_{2}  & =\frac{n_{RE}}{4!}\lim_{\theta\rightarrow0}\left[  \left(
\partial_{\theta}\right)  ^{4}F\right]  +\frac{K_{1}}{3},\label{EqDefK2}\\
K_{3}  & =\frac{n_{RE}}{6!}\lim_{\theta\rightarrow0}\left[  \left(
\partial_{\theta}\right)  ^{6}F\right]  -\frac{2K_{1}}{45}+\frac{2K_{2}}%
{3},\label{EqDefK3}%
\end{eqnarray}
are the magnetic anisotropy constants (MACs) and $\left(  \partial_{\theta
}\right)  ^{m}$ is the $m$-th order partial derivative with respect to $\theta$.

$F$ and the MACs depend on the eigenenergies $E_{n}$ of the 4f
Hamiltonian, Eq.~(\ref{EqHamiltonianGeneral}), by 
Eqs.~(\ref{EqFHelmholtz}) and (\ref{EqDefK1}-\ref{EqDefK3}). For example, when
$J=1$ and in the limit of large exchange ($\left\vert J_{ex}\right\vert
\gg|\vartheta_{l}\left\langle r^{l}\right\rangle A_{l}^{(0)}|$, for $l=2,4,6$)
and low temperatures ($k_{B}T\rightarrow0$),\footnote{Some denote the
magnetic anisotropy energy as $\kappa_{1}\cos^{2}\theta$ instead of $K_{1}%
\sin^{2}\theta$ with $\kappa_{1}=-K_{1}$.}
\begin{equation}
K_{1}=-\frac{3}{2}n_{RE}\theta_{2}\left\langle r^{2}\right\rangle A_{2}%
^{(0)}.\label{EqK1AResult}%
\end{equation}
$K_{1}$ does not depend on the exchange constant~\cite{DefK}.
Equation~(\ref{EqK1AResult}) is consistent with Ref.~\cite{BookSkomski2} and
can be written as $K_{1}/n_{RE}=-(3/2)Q_{2}A_{2}^{(0)}$, where $Q_{2}%
=\theta_{2}\left\langle r^{2}\right\rangle $ is the quadrupolar moment for
$J_{z}=J=1$, a measure of the asphericity of the 4f subshell charge density.

The calculation of general MACs requires the diagonalization of a $(2J+1)$
times $(2J+1)$ matrix. An analytic calculation of $K_{1}$, $K_{2}$ and $K_{3}
$ for arbitrary temperature and exchange constants is tedious, but easily
carried out numerically. 

In the following, we numerically compute the temperature-dependent MACs
induced by electric fields at an \textit{insulator}$|$\textit{metal}
interface, also considering that crystal fields at interfaces may
substantially differ from that in bulk crystals. We assume uniaxial symmetry
and denote the interface crystal field parameters by $\bar{A}_{l}^{(0)}$. An
applied voltage can give rise to locally large electric fields $E_{0}$
normal to a metal$|$insulator interface (along the $z$-axis), which
contributes as $\Delta A_{l}^{(0)}$ with total $\tilde{A}_{l}^{(0)}=\bar
{A}_{l}^{(0)}+\Delta A_{l}^{(0)}$. \begin{figure}[b]
\begin{center}
\includegraphics[width=8.5cm]{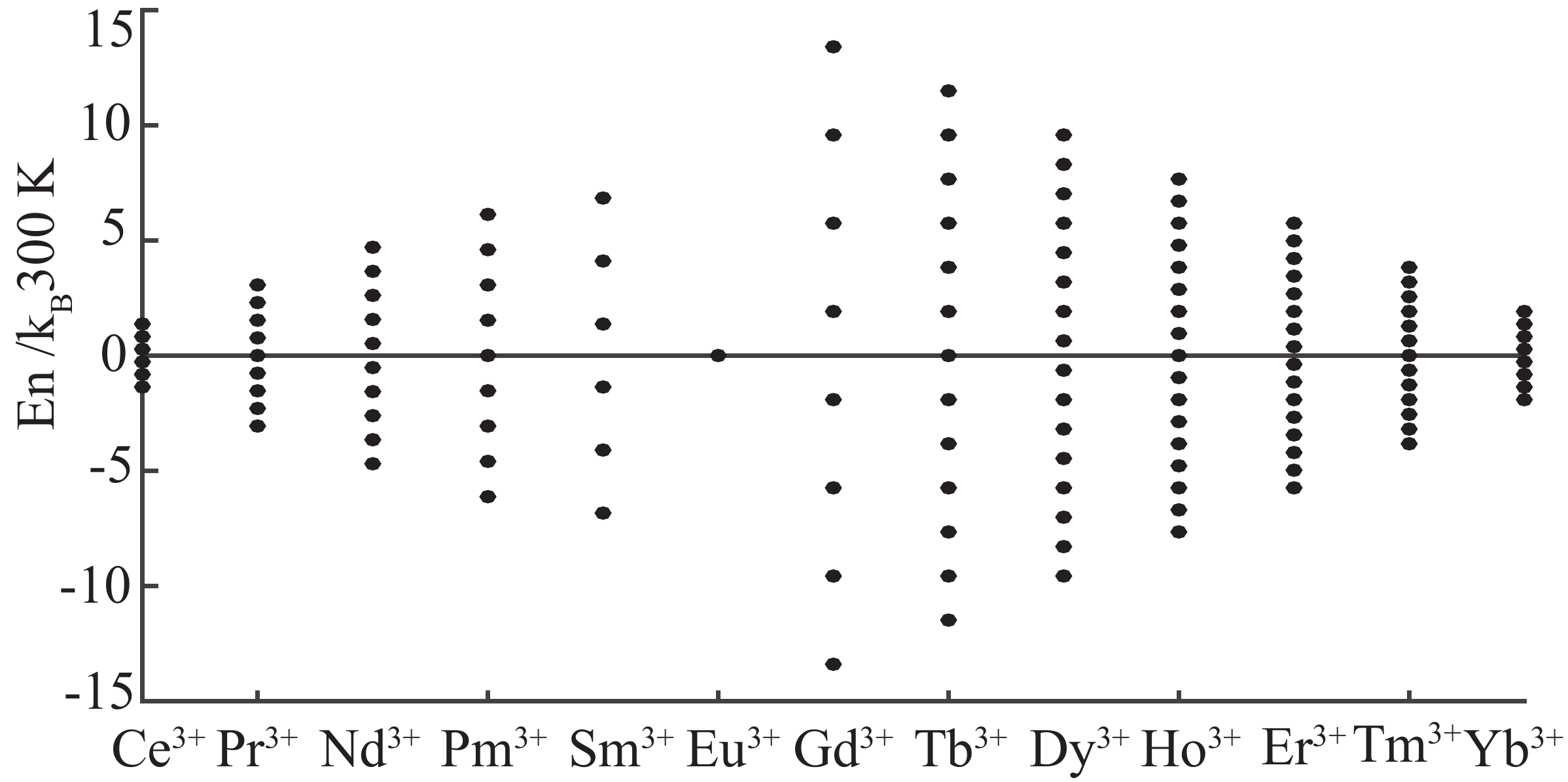}
\end{center}
\caption{Energy spectra of lanthanide local moments at a Co surface at room
temperature, in units of $300$~K$k_{B}$, $J_{ex}=0.1$~eV, $E_{0}=10$~mV/nm, and magnetization along $z $ (i.e.,
$\theta=0$). Crystal-electric field effects are not included. The dots are the calculated eigenvalues
$E_{n}$ of the Hamiltonian~(\ref{EqHamiltonianGeneral}).}%
\label{Fig0}%
\end{figure}

\section{Electric field-dependent magnetic anisotropy}

The applied electric field $E_{0}$ is screened on the scales of the
Thomas-Fermi length $d_{TF}\sim1$~\AA \ on the metal side, so $\mathbf{E}%
=E_{0}e^{-z/d_{TF}}\mathbf{e_{z}}$ for $z>0$, with $z=0$ being the interface
position. Close to $z=0$ and using the expressions from~\ref{SecAppendixA},
\begin{eqnarray}
\Delta A_{2}^{(0)}  & =-\frac{eE_{0}}{6d_{TF}},\\
\Delta A_{4}^{(0)}  & =-\frac{eE_{0}}{840d_{TF}^{3}},\\
\Delta A_{6}^{(0)}  & =-\frac{eE_{0}}{166320d_{TF}^{5}}.
\end{eqnarray}
Therefore, the electric field modifies not only the second-order uniaxial
anisotropy but also higher-order terms. With this set of crystal-field
parameters, we can diagonalize the atomic
Hamiltonian~(\ref{EqHamiltonianGeneral}), evaluate the free
energy~(\ref{EqFHelmholtz}), and the MACs~(\ref{EqDefK1}-\ref{EqDefK3}%
), see~\ref{AppNum}. We plot the
spectra $E_{n}$ with $n \in \{1,\ldots,2J+1\}$ at room temperature in Fig.~\ref{Fig0} in units of
the thermal energy for $J_{ex}=0.1$~eV, $E_{0}=10$~mV/nm, and $\theta=0$. The
exchange interaction dominates the term splittings, while electric-field
effects are small.
\begin{figure}[t]
\begin{center}
\includegraphics[width=8.5cm]{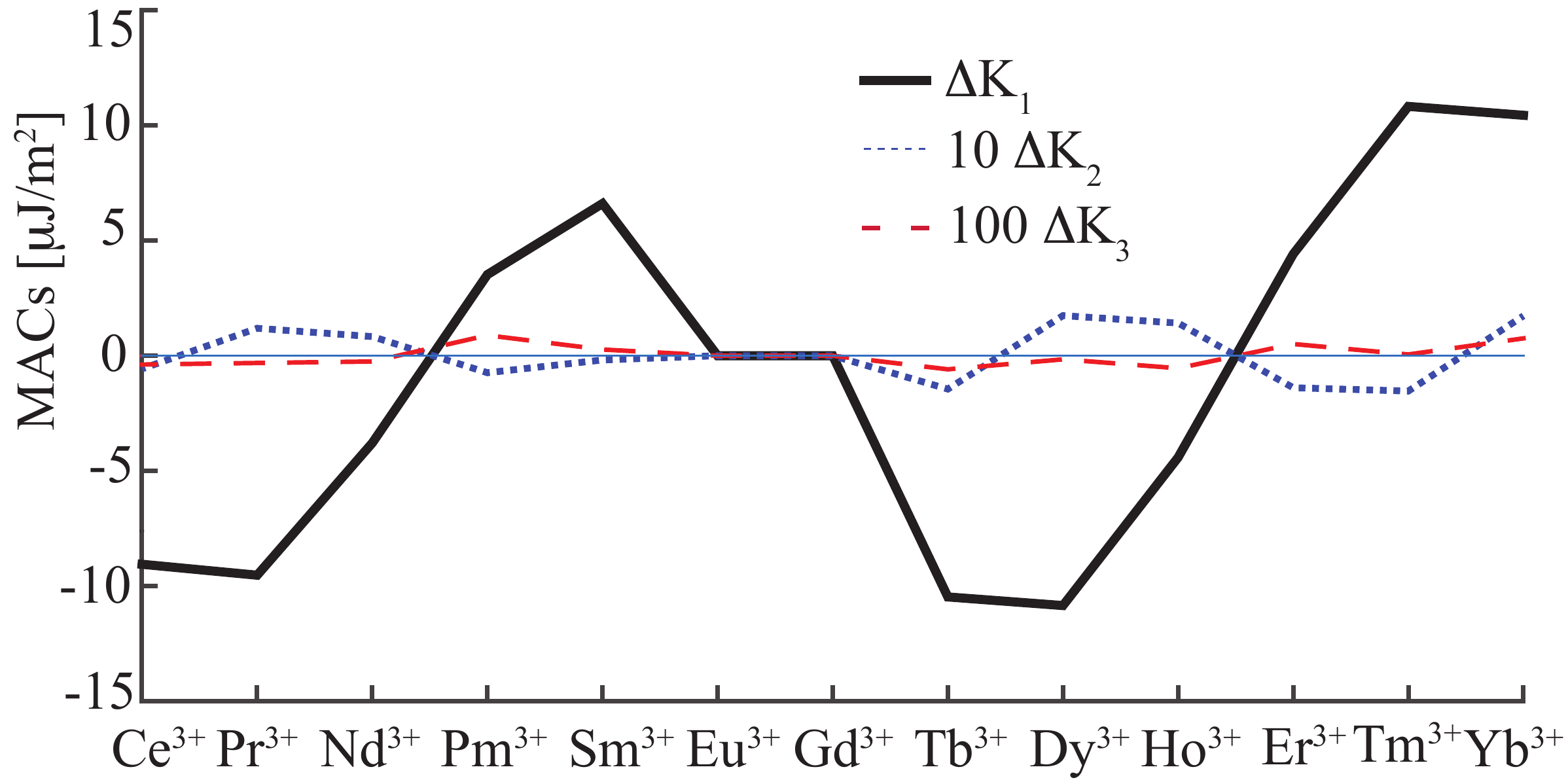}
\end{center}
\caption{Voltage-controlled magnetic anisotropy (MAC) per unit of electric field $E_0$, $\Delta K_{1}/E_0$ (solid line), 
$\Delta K_{2}/E_0$ (square-dashed line) and $\Delta K_{3}/E_0$
(dashed line) of rare earth moments at the surface of Co 
at low temperatures ($T=0.01$ mK). Here we use the density $n_{RE}=1$/nm$^{2}$ and exchange constant $J_{ex}=0.1$ eV. 
For better visibility, $\Delta K_{2}$ and $\Delta K_{3}$ are enlarged
by a factor of 10 and 100, respectively. For an electric field
$E_{0}=10$ mV/nm $=100$ kV/cm, $\Delta K_{1}$ is of the order $\mu$J/m$^2$ for most lanthanides.}%
\label{Fig1}%
\end{figure}

The MACs $\Delta K_{l}$ from $\Delta A_{l}^{(0)}$ are proportional
to the applied electric field $E_{0}$. $\Delta K_{1}$ has a negative slope for the oblate
(pancake-shaped) ions Ce$^{3+}$, Pr$^{3+}$, Nd$^{3+}$, Tb$^{3+}$, Dy$^{3+}$,
and Ho$^{3+}$, and a positive slope for the prolate (cigar-shaped) ions
Pm$^{3+}$, Sm$^{3+}$, Er$^{3+}$, Tm$^{3+}$, and Yb$^{3+}$, consistent with
previous results~\cite{OurVoltage}. Figure~\ref{Fig1} shows the VCMA
contributions of a set of RE atoms at an interface at low temperatures with
$n_{RE}=1$\thinspace nm$^{-2}$, and $J_{ex}=0.1$ eV. We use $d_{TF}=1$~\AA \
and the Co parameters for the magnetization, $\{T_{c},s,p\}=\{1385$%
~K$,0.11,5/2\}$ in Eq. (\ref{EqKuzMin}), assuming that they are not affected
much by the interface. The MACs in units of energy density result from
dividing the surface MACs by the thickness of the magnetic film. For example,
dusting the interface with one Tm$^{+3}$ ion per nm$^{2}$ with a field of $E_{0}%
\sim1$\thinspace$\mathrm{V/nm}=10^{4}\,\mathrm{kV/cm}$ creates an energy volume density of 1
MJ/m$^{3}$ in a 1 nm-thick Co film. Figure~\ref{Fig1} illustrates that the
VCMA of rare earths is governed only by $K_{1}$, while $K_{2}$ and $K_{3}$ are
negligibly small. This hierarchy differs from that of transition metals, where
$K_{1}$ and $K_{2}$ are of the same order of magnitude and partially
compensate each other~\cite{VCMAHigher1,VCMAHigher2}. This difference can be
understood as follows. The $l$-th order MAC divided by the characteristic
electrostatic energy, $eE_{0}d_{TF}$, scales as $\Delta K_{j}/(eE_{0}%
d_{TF})\propto\vartheta_{2j}\langle r^{2j}\rangle/d_{TF}^{2j}$. The 4f
subshell envelope is nearly ellipsoidal, which is accounted for by the
hierarchy of the projections constants $\left\vert \vartheta_{2}\right\vert
\ll\left\vert \vartheta_{4}\right\vert \ll\left\vert \vartheta_{6}\right\vert
$. The transition metal 3d shells are more polarizable and can be more easily
deformed by the crystal fields than the lanthanides. A consequence is that the
quadrupole contribution of the voltage-controlled anisotropy $\Delta K_{1}$ of
rare earths is much larger than $\Delta K_{2}$ and $\Delta K_{3}$.

\begin{figure}[t]
\begin{center}
\includegraphics[width=8.5cm]{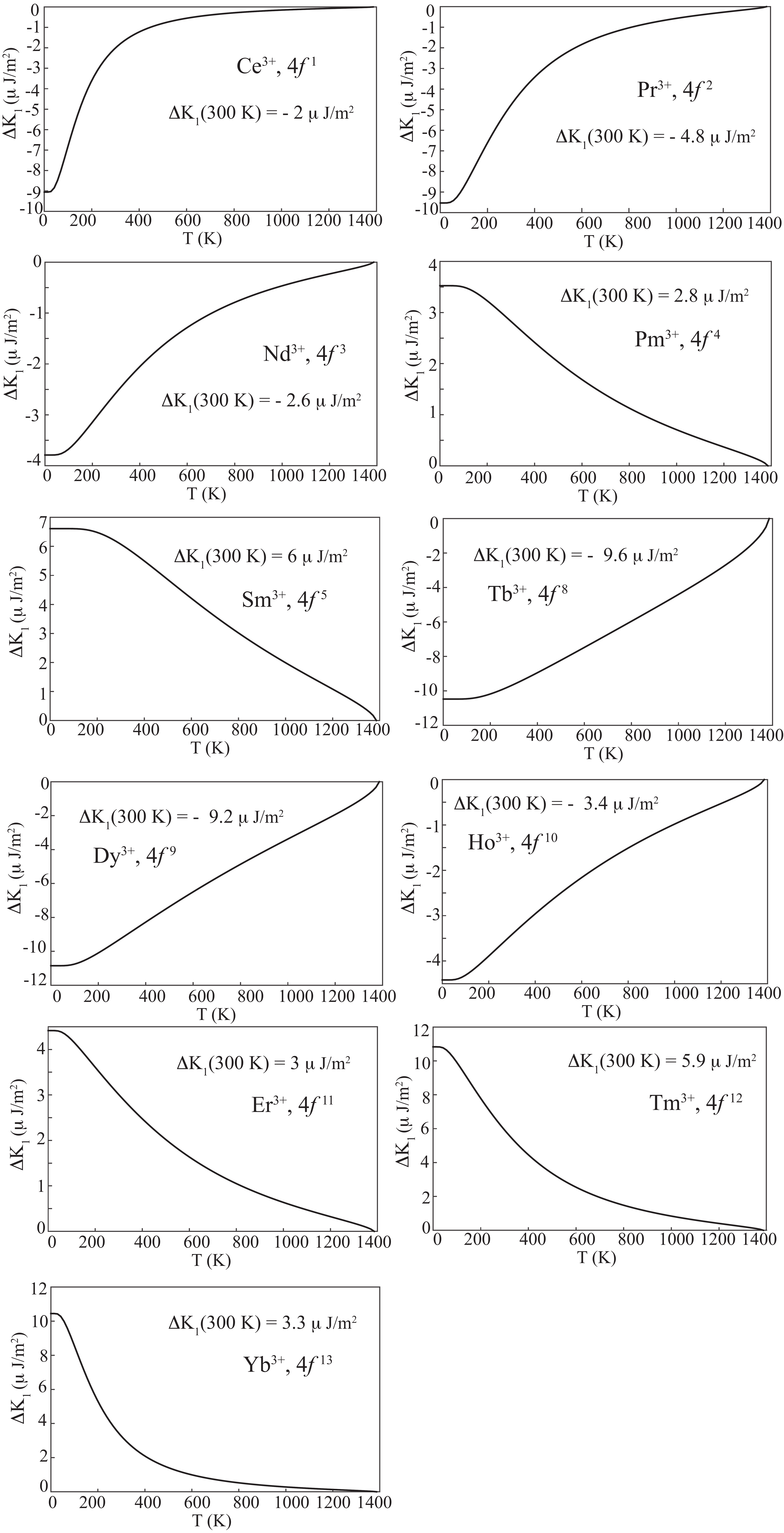}
\end{center}
\caption{Magnetic anisotropy constants per unit of electric field, $\Delta K_{1}/E_0$, as a function of
temperature for a rare-earth density $n_{4f}=1$/nm$^{2}$
at a Co surface.}%
\label{Fig2}%
\end{figure}

The temperature dependence of rare-earth magnetic anisotropies in bulk
materials has been extensively
studied~\cite{BookSkomski1,BookSkomski2,KuzMin2}. Here we calculate the
temperature dependence of the VCMA-efficiency for rare-earth atoms at an interface
between a non-magnetic insulator (such as MgO) and a magnetic metal, such as
Fe or Co. Figure~\ref{Fig2} illustrates the temperature-dependence of $K_{1}/E_0$
for all lanthanides with a finite orbital momentum in the temperature range
0~K~$\leq T\leq$~1400 K~for $n_{RE}=1$~nm$^{-2}$. $K_{1}$ at room temperature,
$T=300$ K, is specified inside each graph. The efficiency at room-temperature is
largest for Tb$^{3+}$ and Dy$^{3+}$ with $\Delta K_{1}/E_0=-960$ fJ/Vm and $\Delta K_{1}/E_0=-910$ fJ/Vm, respectively. For an applied field of $E_0=10$mV/nm, the corresponding VCMA values of Tb$^{3+}$ and Dy$^{3+}$ are $\Delta K_{1}=-9.6\,\mathrm{\mu}$J/m$^{2}$ and $\Delta K_{1}=-9.1\,\mathrm{\mu}$J/m$^{2}$.

\begin{figure}[t]
\begin{center}
\includegraphics[width=8.5cm]{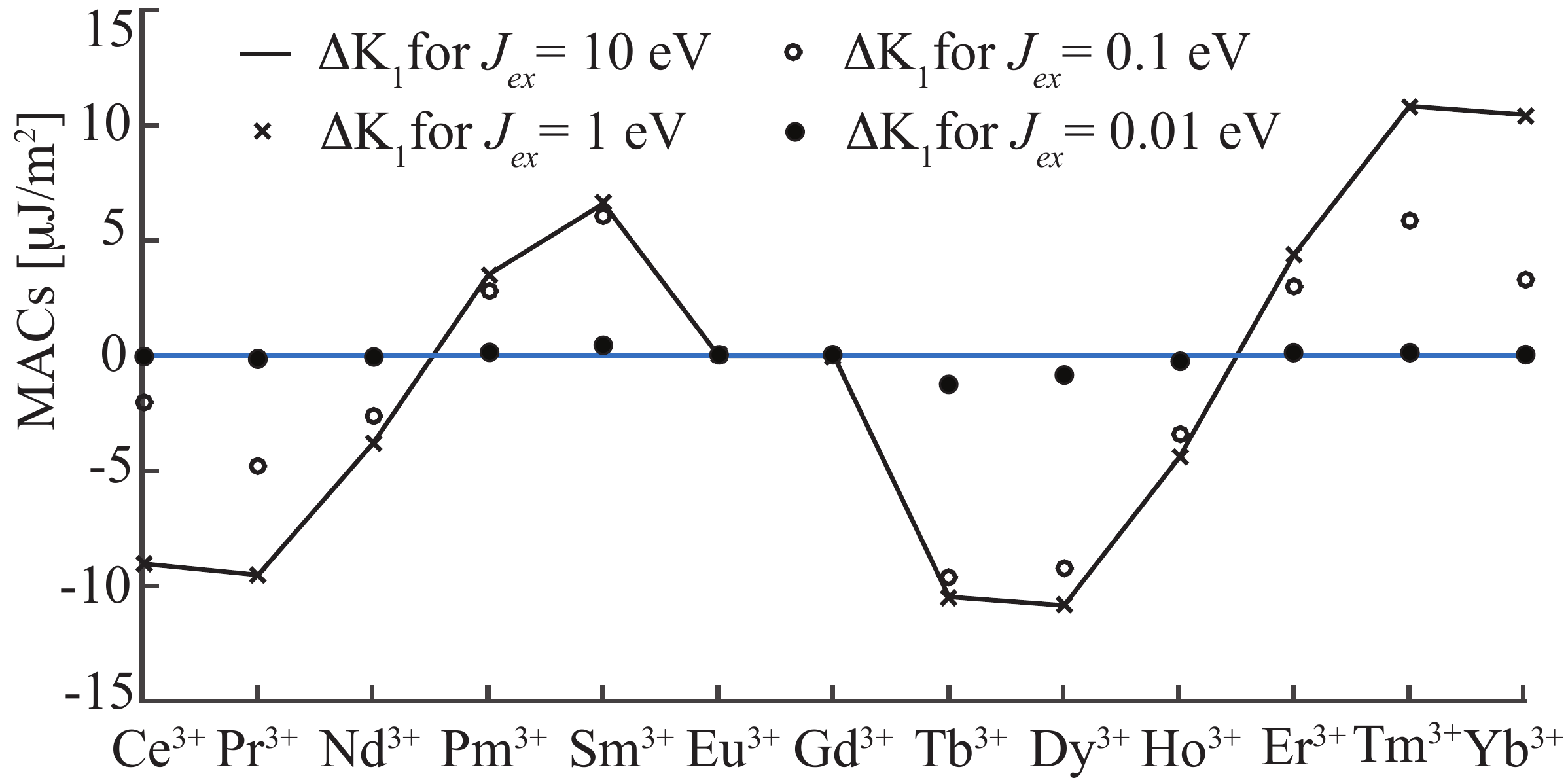}
\end{center}
\caption{Magnetic anisotropy constant per unit of electric field, $\Delta K_{1}/E_0$, for exchange constants $J_{ex}=10$ eV (solid
line), $J_{ex}=1$ eV (crosses), $J_{ex}=0.1$ eV (open circles), and
$J_{ex}=0.01$ eV (full circles). This graph uses $T=300$ K
and a density $n_{4f}=1$/nm$^{2}$ at a Co surface. The thin horizontal line
$\Delta K_{1}/E_0=0$ is just for visual guidance.}%
\label{Fig3}%
\end{figure}

In the absence of exchange coupling between the 4f angular momentum
($\mathbf{J}$) and the magnetization ($\mathbf{m}$), REs do not contribute to
the anisotropy, so the VCMA strength vanishes for $J_{ex}\rightarrow0$. This
tendency is shown in Fig.~\ref{Fig3} for $0.01$~$\mathrm{eV}\leq J_{ex}\leq10$
eV at $T=300$~K. Results are not very sensitive to the value of typical
exchange constants, $0.1$~$\mathrm{eV}\leq J_{ex}<1$ eV, as long as they are
larger than the anisotropy induced by the crystal fields or applied voltages
($\sim0.01$~eV~\cite{BookSkomski1}).

\section{Intrinsic interface magnetic anisotropy}

\label{Intrinsic} The intrinsic (i.e., at zero applied electric field) magnetic anisotropy at the interface cannot be calculated accurately in an easy way. Simple approaches, such as the
point-charge model, are not adequate for metals due to the efficient screening
by conduction electrons~\cite{SkomskiScreenedPointCharge}. The screened-charge
model of metals~\cite{SkomskiScreenedPointCharge} can
characterize interfacial anisotropies in metallic
multilayers~\cite{SkomskiScreenedPMultilayer}. However, this model is not
valid for metal$|$insulator interfaces with a nearly discontinuous conduction electron density.

\begin{figure}[b]
\begin{center}
\includegraphics[width=4.5cm]{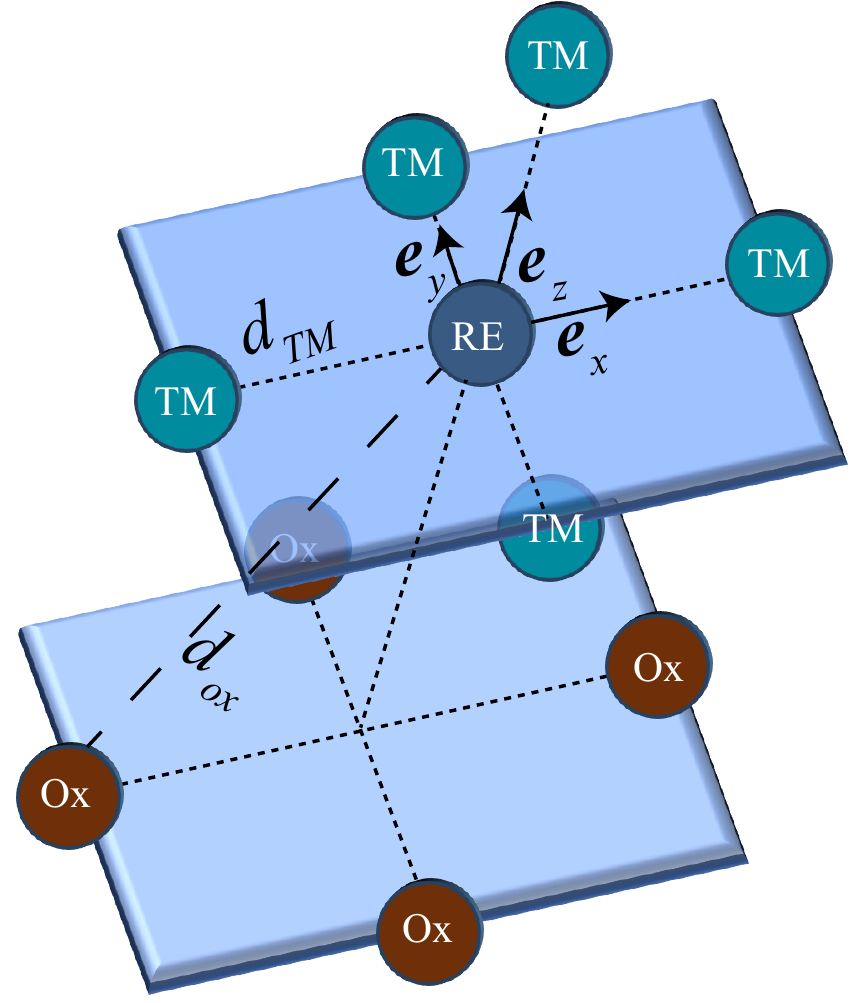}
\end{center}
\caption{Sketch of the ligands of a RE atom at a metal$|$insulator
interface.}%
\label{Fig4}%
\end{figure}

Here we estimate the order of magnitude of the intrinsic interfacial RE
magnetic anisotropy by the model of a local moment in a metal at the origin
surrounded by four oxygen atoms with Cartesian coordinates $(\pm
d_{ox},0,-d_{ox})/\sqrt{2}$ and $(0,\pm d_{ox},-d_{ox})/\sqrt{2}$ and five
transition-metal atoms (such as Co or Fe) at positions $(\pm d_{TM},0,0)$,
$(0,\pm d_{TM},0)$ and $(0,0,d_{TM})$, as shown in Fig.~\ref{Fig4}. The
uniaxial crystal-field
parameter~\cite{BookSkomski1,BookSkomski2,SkomskiScreenedPointCharge} reads
\begin{equation*}
\bar{A}_{2}^{(0)}=\sum_{j}\frac{A_{j}^{\prime}}{2}\left(  3\cos^{2}\theta
_{j}-1\right)  ,
\end{equation*}
where $j$ labels the ligand, $\cos\theta_{j}$ is the $z$-component of the
$j$-th site position ($\mathbf{r}_{j}$), and $A_{j}^{\prime}$ depends on the
distance $d_{j}=|\mathbf{r}_{j}|$
\begin{equation*}
A_{j}^{\prime}\left(  d_{j}\right)  =-\frac{eQ_{j}}{4\pi\varepsilon_{0}}%
\frac{e^{-d_{j}/d_{TF}}}{2d_{j}^{3}}\left[  1+\frac{d_{j}}{d_{TF}}+\frac{1}%
{3}\left(  \frac{d_{j}}{d_{TF}}\right)  ^{2}\right]  ,\label{EqIntrinsicCFP}%
\end{equation*}
where $\varepsilon_{0}$ is the vacuum permittivity. We adopt the
screened charges~\cite{SkomskiScreenedPointCharge} approach for $Q_{j}=Q_{j}\left(d_{TF}\right)$.  For $d_{ox}=6
$~\AA , $d_{TM}=5$~\AA , and $d_{TF}=1$~\AA , and $A_{j}^{\prime}/k_{B}$
for iron and oxygen of the order of magnitude of
$100\,\mathrm{K}a_{0}^{-2}$ and $200$\thinspace
$\mathrm{K}a_{0}^{-2}$, respectively, $\bar{A}_{2}^{(0)}\sim3\times10^{18}%
$~eV/m$^{2}$ is of the same order as that produced by an electric field of
$E_{0}=-1.8$\thinspace V/nm.

For the present interface model, oblate (prolate) ions with
$\vartheta_{2}<0$ ($\vartheta_{2}>0$) favor a perpendicular (in-plane)
magnetization. Doping a transition-metal layer with oblate rare-earth ions
enhances the perpendicular interface anisotropy, which is important for
spin transfer torque magnetic random-access memories (STT-MRAM), but also implies need for higher voltages to achieve VCMA-induced
magnetization switching. In
transition metals, on the other hand, the intrinsic magnetic anisotropies are small, and electric-field
effects easily dominate. A quantitative description of the intrinsic
interface rare-earth anisotropy as a function of interface structure and
morphology requires first-principles calculations.

\section{Conclusions and discussion}

We studied the temperature-dependent \textit{voltage-controlled magnetic anisotropy} of
rare-earth atoms at a magnetic metal$|$non-magnetic insulator interface. Our findings differ from the conventional wisdom based on
transition metals. In rare earths, only the lowest-order uniaxial constant can
be efficiently modulated by a voltage because of the small 4f radius and rigid
ellipsoidal shape of the 4f shell electron density. To leading order, the magnetic
anisotropy constants change linearly with the applied electric field, with a
negative slope for the oblate (pancake-like) Ce$^{3+}$, Pr$^{3+}$, Nd$^{3+}$,
Tb$^{3+}$, Dy$^{3+}$, and Ho$^{3+}$, and a positive one for the prolate
Pm$^{3+}$, Sm$^{3+}$, Er$^{3+}$, Tm$^{3+}$, and Yb$^{3+}$ moments. Rare
earths at an interface also contribute to the intrinsic (i.e., independent of the applied electric field) magnetic anisotropy,
the oblate (prolate) ones favoring a perpendicular (in-plane) equilibrium magnetization.

Our model assumes metallic screening, i.e., a drop of the electric field over
atomic distances at the interface which hosts the rare earth moments. This
assumption might break down at non-ideal interfaces, so it should be confirmed
by experimental or ab initio methods.

Nevertheless, we are confident about substantial effects at room temperature
for even low densities of RE atoms ($\sim1$/nm$^{2}$). Since the electric
field is strongly enhanced at metal$|$insulator interfaces, bulk doping of a
magnet with rare earths is not efficient. Still, the dusting of the
interface between a tunnel barrier and a transition metal thin film can
significantly enhance the switching efficiency of voltage-controlled tunnel junctions.

\subsection*{Acknowledgments}

This research was supported by JSPS KAKENHI Grant No. 19H006450, Postdoctorado
FONDECYT 2019 Folio 3190030, and Financiamento Basal para Centros Cientificos
de Excelencia FB0807.

\appendix

\section{Expansion into spherical harmonics}

\label{SecAppendixA} We focus on the axially symmetric potentials, $-eV\left(
\mathbf{r}\right)  $, that we decompose into the Spherical harmonics
$Y_{l}^{0}\left(  \theta\right)  $
\begin{equation}
Y_{l}^{0}\left(  \theta\right)  =\sqrt{\frac{2l+1}{4\pi}}P_{l}\left(
\cos\theta\right)  ,
\end{equation}
where $l=2,4,6$ and $P_{l}$ is the $l$-th Legendre polynomial,
\begin{eqnarray}
P_{2}(x)  & =\frac{1}{2}\left(  3x^{2}-1\right)  ,\\
P_{4}(x)  & =\frac{1}{8}\left(  35x^{4}-30x^{2}+3\right)  ,\\
P_{6}(x)  & =\frac{1}{16}\left(  231x^{6}-315x^{4}+105x^{2}-5\right)  .
\end{eqnarray}
leading to a multipolar expansion up to the 6th order in $r$
\begin{eqnarray}
-eV\left(  \mathbf{r}\right)   & =4\sqrt{\frac{\pi}{5}}Y_{2}^{0}(\theta
)r^{2}A_{2}^{(0)}+16\sqrt{\frac{\pi}{9}}Y_{4}^{0}(\theta)r^{4}A_{4}%
^{(0)}\nonumber\\
& +32\sqrt{\frac{\pi}{13}}Y_{6}^{0}(\theta)r^{6}A_{6}^{(0)}+c_{0}\left(
\theta\right)  ,
\end{eqnarray}
where $c_{0}\left(  \theta\right)  $ collects the odd terms in $z$
(dipolar-like contributions) that do not interact with a nearly ellipsoidal 4f
subshell (i.e., the fields are not large enough to polarize/asymmetrize the
subshell). Using the orthonormality of spherical harmonic functions, $\int%
_{0}^{\pi}d\theta\sin\theta\int_{0}^{2\pi}d\phi\left[  Y_{l^{\prime}%
}^{m^{\prime}}(\theta,\phi)\right]  ^{\ast}Y_{l}^{m}(\theta,\phi
)=\delta_{l,l^{\prime}}\delta_{m,m^{\prime}}$, one gets
\begin{eqnarray}
A_{2}^{(0)}  & =\frac{-e}{4r^{2}}\sqrt{\frac{5}{\pi}}\int_{0}^{\pi}d\theta
\sin\theta\int_{0}^{2\pi}d\phi Y_{2}^{0}(\theta)V\left(  \mathbf{r}\right)
,\\
A_{4}^{(0)}  & =\frac{-e}{16r^{4}}\sqrt{\frac{9}{\pi}}\int_{0}^{\pi}%
d\theta\sin\theta\int_{0}^{2\pi}d\phi Y_{4}^{0}(\theta)V\left(  \mathbf{r}%
\right)  ,\\
A_{6}^{(0)}  & =\frac{-e}{32r^{6}}\sqrt{\frac{13}{\pi}}\int_{0}^{\pi}%
d\theta\sin\theta\int_{0}^{2\pi}d\phi Y_{6}^{0}(\theta)V\left(  \mathbf{r}%
\right)  .
\end{eqnarray}
The single-electron 4f wave functions have principal and orbital quantum
numbers $n=4$ and $l=3$, respectively. Consequently, the 4f charge
distribution has non-vanishing multipoles up to the $2l=6$-th order.

\section{Numerical Details}

\label{AppNum}

For convenience, we introduce dimensionless parameters for the MACs
$k_{i}\equiv\Delta K_{i}/(n_{RE}J_{ex})$, reciprocal thermal energy
$\bar{\beta}\equiv J_{ex}\beta$, and crystal-field parameters $a_{l}%
\equiv\vartheta_{l}\left\langle r^{l}\right\rangle A_{l}^{(0)}/J_{ex}$. The
reduced Helmholtz free energy $F/J_{ex}=-\bar{\beta}^{-1}\ln\sum
e^{-\bar{\beta}\epsilon_{n}}$, where $\epsilon_{n}\equiv E_{n}/J_{ex}$ is the
$n$-th eigenvalue of the dimensionless 4f Hamiltonian ${H}/{J_{ex}}$. We
approximate the radial 4f wave function by a Slater-type orbital, $R(r)\propto
r^{3}e^{-r/a}$, with $a=0.133$ \AA , such that the mean value $\left\langle
r\right\rangle =0.6$ \AA ~\cite{BookJensen}. Then, $\left\langle
r^{2}\right\rangle =0.4$ \AA $^{2}$, $\left\langle r^{4}\right\rangle =0.24$
\AA $^{4}$, and $\left\langle r^{6}\right\rangle =0.19$ \AA $^{6}$. We use the
$T_{C}$, $s$ and $p$ values of Co for the temperature dependence of the host
magnetization~\cite{KuzMin1}. The derivatives of Eqs.~(\ref{EqDefK1}%
),~(\ref{EqDefK2}), and~(\ref{EqDefK3}) are discretized using central schemes
of order $\Delta\theta^{2}$, with the $\theta$ step-size $\Delta\theta=0.1$. A
finer grid $\Delta\theta=0.05$ and $\Delta\theta=0.01$ leads to the same results.


\bigskip

\begin{thebibliography}{99}                                                                                               %
\bibitem {STT}Ralph D C and Stiles M D 2008 Spin transfer torques \textit{J.
Magn. Magn. Mater.} \textbf{320} 1190

\bibitem {TmIG0}Avci C O, Quindeau A, Pai C-F, Mann M, Caretta L, A. Tang A S,
Onbasli M C, Ross C A and Beach G S D 2017 Current-induced switching in a
magnetic insulator \textit{Nat. Mater.} \textbf{16} 309

\bibitem {SpinCaloritronics}Bauer G E W, Saitoh E and van Wees B J 2012 Spin
caloritronics \textit{Nat. Mater.} \textbf{11} 391

\bibitem {OtaniSpinConv}Otani Y, Shiraishi M, Oiwa A, Saitoh E and Murakami S
2017 Spin conversion on the nanoscale \textit{Nat. Phys.} \textbf{13} 829

\bibitem {MechanicalWaves1}Matsuo M, Ieda J, Saitoh E and Maekawa S 2011 Spin
current generation due to mechanical rotation in the presence of impurity
scattering \textit{Appl. Phys. Lett.} \textbf{98} 242501

\bibitem {MechanicalWaves2}Kobayashi D, Yoshikawa T, Matsuo M, Iguchi R,
Maekawa S, Saitoh E and Nozaki Y 2017 Spin Current Generation Using a Surface
Acoustic Wave Generated via Spin-Rotation Coupling \textit{Phys. Rev. Lett.}
\textbf{119} 077202

\bibitem {Cavitronics1}Zhang X, Zou C-L, Jiang L and Tang H X 2014 Strongly
Coupled Magnons and Cavity Microwave Photons \textit{Phys. Rev. Lett.}
\textbf{113} 156401

\bibitem {Cavitronics2}Yu W, Wang J, Yuan H Y and Xiao J 2019 Prediction of
Attractive Level Crossing via a Dissipative Mode \textit{Phys. Rev. Lett.}
\textbf{123} 227201

\bibitem {Suzuki2011}Suzuki Y, Kubota H, Tulapurkar A and Nozaki T 2011 Spin
control by application of electric current and voltage in FeCoMgO junctions
\textit{Phil. Trans. R. Soc. A} \textbf{369} 3658

\bibitem {Nozaki2012}Nozaki T, Shiota Y, Miwa S, Murakami S, Bonell F,
Ishibashi S, Kubota H, Yakushiji K, Saruya T, Fukushima A, Yuasa S, Shinjo T
and Suzuki Y 2012 Electric-field-induced ferromagnetic resonance excitation in
an ultrathin ferromagnetic metal layer \textit{Nat. Phys.} \textbf{8} 491

\bibitem {ParametricVCMA}Verba R, Tiberkevich V, Krivorotov I and Slavin A
2014 Parametric Excitation of Spin Waves by Voltage-Controlled Magnetic
Anisotropy \textit{Phys. Rev. Appl.} \textbf{1} 044006

\bibitem{GeneralReferenceEField}
Matsukura F, Tokura Y and Ohno H 2015
Control of magnetism by electric fields
\textit{Nat. Nano.} \textbf{10} 209

\bibitem{GeneralReferenceEField2}
Song C, Cui B, Li F, Zhou X and Pan F 2017
Recent progress in voltage control of magnetism: Materials, mechanisms, and performance
\textit{Prog. Mater.} \textbf{87} 33

\bibitem {VCMARecentReview}Nozaki T, Yamamoto T, Miwa S, Tsujikawa M, Shirai
M, Yuasa S and Suzuki Y 2019 Recent Progress in the Voltage-Controlled
Magnetic Anisotropy Effect and the Challenges Faced in Developing
Voltage-Torque MRAM \textit{Micromachines} \textbf{10} 327

\bibitem {Shiota2009}Shiota Y, Maruyama T, Nozaki T, Shinjo T, Shiraishi M and
Suzuki Y 2009 Voltage-Assisted Magnetization Switching in Ultrathin Fe$_{80}%
$Co$_{20}$ Alloy Layers \textit{Appl. Phys. Express} \textbf{2} 063001

\bibitem{Assisted2}
Shiota Y, Nozaki T, Bonell F, Murakami S, Shinjo T and Suzuki T
2012 Induction of coherent magnetization switching in a few atomic layers of FeCo using voltage pulses
\textit{Nat. Mater.} \textbf{11} 39

\bibitem{Assisted}
Wang W-G, Li M, Hageman S and Chien CL 2012
Electric-field-assisted switching in magnetic tunnel junctions \textit{Nat. Mater.} \textbf{11} 64

\bibitem {Kanai2012}Kanai S, Yamanouchi M, Ikeda S, Nakatani Y, Matsukura F
and Ohno H 2012 Electric field-induced magnetization reversal in a
perpendicular-anisotropy CoFeB-MgO magnetic tunnel junction \textit{Appl.
Phys. Lett.} \textbf{101} 122403

\bibitem{SwitchingNp1}
Kanai S, Matsukura F and Ohno H 2016
Electric-field-induced magnetization switching in CoFeB/MgO magnetic tunnel junctions with high junction resistance
\textit{Appl. Phys. Lett.} \textbf{108}, 192406

\bibitem{SwitchingNp2}
Grezes C, Ebrahimi F, Alzate J G, Cai X, Katine J A, Langer J, Ocker B, Amiri P K and Wang K L 2016
Ultra-low switching energy and scaling in electric-field-controlled nanoscale magnetic tunnel junctions with high resistance-area product
\textit{Appl. Phys. Lett.} \textbf{108}, 012403

\bibitem {Zhu2012}Zhu J, Katine J A, Rowlands G E, Chen Y J, Duan Z, Alzate J
G, Upadhyaya P, Langer J, Amiri P K, Wang K L and Krivorotov I N 2012
Voltage-Induced Ferromagnetic Resonance in Magnetic Tunnel Junctions
\textit{Phys. Rev. Lett. }\textbf{108} 197203

\bibitem {Interfaces2}Wen Z, Sukegawa H, Seki T, Kubota T, Takanashi K and
Mitani S 2017 Voltage control of magnetic anisotropy in epitaxial Ru/Co2FeAl/
MgO heterostructures \textit{Sci. Rep.} \textbf{7} 45026

\bibitem {ThermalStability}Shiota Y, Nozaki T, Tamaru S, Yakushiji K, Kubota
H, Fukushima A, Yuasa S and Suzuki Y 2017 Reduction in write error rate of
voltage-driven dynamic magnetization switching by improving thermal stability
factor \textit{Appl. Phys. Lett.} \textbf{111} 022408

\bibitem {VCMAHigher1}Sugihara A, Nozaki T, Kubota H, Imamura H, Fukushima A,
Yakushiji K and Yuasa S 2019 Evaluation of higher order magnetic anisotropy in
a perpendicularly magnetized epitaxial ultrathin Fe layer and its applied
voltage dependence \textit{Jpn. J. Appl. Phys.} \textbf{58} 090905

\bibitem {VCMAHigher2}Sugihara A, Spiesser A, Nozaki T, Kubota H, Imamura H,
Fukushima A, Yakushiji K and Yuasa S 2020 Temperature dependence of
higher-order magnetic anisotropy constants and voltage-controlled magnetic
anisotropy effect in a Cr/Fe/MgO junction \textit{Jpn. J. Appl. Phys.}
\textbf{59} 010901

\bibitem {Gerhard2010}Gerhard L, Yamada T K, Balashov T, Tak\'acs A F,
Wesselink R J H, D\"ane M, Fechner M, Ostanin S, Ernst A, Mertig I and
Wulfhekel W 2010 Magnetoelectric coupling at metal surfaces \textit{Nat.
Nano.} \textbf{5} 792

\bibitem {Yamada2011}Yamada Y, Ueno K, Fukumura T, Yuan H T, Shimotani H,
Iwasa Y, Gu L, Tsukimoto S, Ikuhara Y and Kawasaki M 2011 Electrically Induced
Ferromagnetism at Room Temperature in Cobalt-Doped Titanium Dioxide
\textit{Science} \textbf{332} 1065

\bibitem {Sekine2016}Sekine A and Chiba T 2016 Electric-field-induced spin
resonance in antiferromagnetic insulators: Inverse process of the dynamical
chiral magnetic effect \textit{Phys. Rev. B} \textbf{93} 220403(R)

\bibitem {Platinum}Miwa S, Suzuki M, Tsujikawa M, Matsuda K, Nozaki T, Tanaka
K, Tsukahara T, Nawaoka K, Goto M, Kotani Y, Ohkubo T, Bonell F, Tamura E,
Hono K, Nakamura T, Shirai M, Yuasa S and Suzuki Y 2017 Voltage controlled
interfacial magnetism through platinum orbits \textit{Nat. Comm.} \textbf{8} 15848

\bibitem {PtProximityMag}Liang L, Shan J, Chen Q H, Lu J M, Blake G R, Palstra
T T M, Bauer G E W, van Wees B J and Ye J T 2018 Gate-controlled
magnetoresistance of a paramagnetic-insulator|platinum interface \textit{Phys.
Rev. B} \textbf{98} 134402

\bibitem {ChibaTanaka}Takiguchi K, Anh L D, Chiba T, Koyama T, Chiba D and
Tanaka M 2019 Giant gate-controlled proximity magnetoresistance in
semiconductor-based ferromagnetic$\vert$non-magnetic bilayers \textit{Nat.
Phys.} \textbf{15} 1134

\bibitem {SMMWernsdorfer}Thiele S, Balestro F, Ballou R, Klyatskaya S, Ruben M
and Wernsdorfer W 2014 Electrically driven nuclear spin resonance in
single-molecule magnets \textit{Science} \textbf{344} 6188

\bibitem {You2005}You C-Y and Suzuki Y 2005 Tunable interlayer exchange
coupling energy by modification of Schottky barrier potentials \textit{J.
Magn. Magn. Mater.} \textbf{293} 774

\bibitem {Haney2009a}Haney P M, Heiliger C and Stiles M D 2009 Bias dependence
of magnetic exchange interactions: Application to interlayer exchange coupling
in spin valves \textit{Phys. Rev. B} \textbf{79} 054405

\bibitem {Tang2009}Tang Y-H, Kioussis N, Kalitsov A, Butler W H and Car R 2009
Controlling the Nonequilibrium Interlayer Exchange Coupling in Asymmetric
Magnetic Tunnel Junctions \textit{Phys. Rev. Lett.} \textbf{103} 057206

\bibitem {New2017}Newhouse-Illige T, Liu Y, Xu M, Hickey D R, Kundu A, Almasi
H, Bi C, Wang X, Freeland J W, Keavney D J, Sun C J, Xu Y H, Rosales M, Cheng
X M, Zhang S, Mkhoyan K A and Wang W G 2017 Voltage-controlled interlayer
coupling in perpendicularly magnetized magnetic tunnel junctions \textit{Nat.
Comm.} \textbf{8} 15232

\bibitem {RKKYMultilayers1}Yang Q, Wang L, Zhou Z, Wang L, Zhang Y, Zhao S,
Dong G, Cheng Y, Min T, Hu Z, Chen W, Xia K and Liu M 2018 Ionic liquid gating
control of RKKY interaction in FeCoB/Ru/FeCoB and (Pt/Co)2/Ru/(Co/Pt)2
multilayers \textit{Nat. Comm.} \textbf{9} 991

\bibitem {RKKYImpurities}Leon A O, d'Albuquerque e Castro J, Retamal J~C,
Cahaya A B and Altbir D 2019 Manipulation of the RKKY exchange by voltages
\textit{Phys. Rev. B} \textbf{100} 014403

\bibitem {BookSkomski1}Skomski R 2008 \textit{Simple Models of Magnetism}
(Oxford University Press)

\bibitem {BookSkomski2}Skomski R and Coey J M D 1999 \textit{Permanent
Magnetism} (Institute of Physics Publishing)

\bibitem {OurVoltage}Leon A O, Cahaya A B and Bauer G E W 2018, Voltage
Control of Rare-Earth Magnetic Moments at the Magnetic-Insulator-Metal
Interface \textit{Phys. Rev. Lett.} \textbf{120} 027201

\bibitem {BookJensen}Jensen J and Mackintosh A R 1991 \textit{Rare Earth
Magnetism} (Oxford: Clarendon Press)

\bibitem {CoeysBook}Coey J M D (ed) 1996 \textit{Rare-earth Iron Permanent
Magnets} (Oxford: Clarendon Press)

\bibitem {StevensOp1}Stevens K W H 1951 Matrix Elements and Operator
Equivalents Connected with the Magnetic Properties of Rare Earth Ions
\textit{Proc. Phys. Soc.} \textbf{65} 209-215

\bibitem {StevensOp2}Hutchings M T 1964 Point-Charge Calculations of Energy
Levels of Magnetic Ions in Crystalline Electric Fields \textit{Solid State
Phys.} \textbf{16} C 227-273

\bibitem {StevensOp3}Smith D and Thornley J H M 1966 The use of "operator
equivalents" \textit{Proc. Phys. Soc. }\textbf{89} 779

\bibitem {KuzMin1}Kuz'min M D 2005 Shape of Temperature Dependence of
Spontaneous Magnetization of Ferromagnets: Quantitative Analysis \textit{Phys.
Rev. Lett.} \textbf{94} 107204

\bibitem {KuzMin2}Miura D and Sakuma A 2019 Non-collinearity Effects on
Magnetocrystalline Anisotropy for R2Fe14B Magnets \textit{J. Phys. Soc. Jpn.}
\textbf{88} 044804

\bibitem {DefK}Daisuke M, Sasaki R and Sakuma A 2015 Direct expressions for
magnetic anisotropy constants \textit{Appl. Phys. Express }\textbf{8} 113003

\bibitem {SkomskiScreenedPointCharge}Skomski R 1994 The screened-charge model
of crystal-field interaction \textit{Philos. Mag. B} \textbf{70} 175

\bibitem {SkomskiScreenedPMultilayer}Skomski R, Brennan S and Coey J M D 1995
Crystal-Field Screening in Metallic Multilayers \textit{J. Mag. Soc. Japan}
\textbf{19} 49
\end{thebibliography}
\end{document}